# AN ALTERNATIVE APPROACH TO THE PHONON THEORY OF LIQUIDS: AN ANALYTICAL STUDY OF FRENKEL FREQUENCY AND HEAT CAPACITY AS A FUNCTION OF PRESSURE AND TEMPERATURE

M. Y. Esmer[1, a)] and Bahtiyar A. Mamedov[1]

[1]Department of Physics, Faculty of Arts and Sciences, Gaziosmanpaşa University, Tokat, Türkiye

[a)] Authors to whom correspondence should be addressed: mylmzesmer@hotmail.com

## ABSTRACT

Based on the Maxwell relationship and experimental viscosity data, the phonon theory of liquids can provide a temperature-dependent description of liquid heat capacity that is consistent with experimental data. However, since liquid heat capacity also varies with pressure, we present an alternative approach that can be used to calculate the Frenkel frequency in terms of temperature and pressure by applying the concept of chemical potential under the assumption of a diffusive equilibrium. Using this derived Frenkel frequency, we formulate an analytical expression for the liquid heat capacity in terms of both temperature and pressure, without the need for viscosity data, which is consistent with predictions from the phonon theory of liquids. Our model is tested by comparing the calculated heat liquid capacity with experimental data for four noble liquids at various pressures and temperatures, and good agreement is found. Finally, based on our findings, we propose analytical expressions for the Frenkel line and viscosity as a function of pressure and temperature, and discuss the key details and implications of our approach.

## I. INTRODUCTION

Calculation of the thermodynamic properties of liquids such as energy and heat capacity poses a theoretical challenge, due to their strong, system-specific interactions, in contrast to theories of solids and gases [1]. This issue does not arise for strongly-interacting solids due to the small oscillations around equilibrium points, which permit expansion of the potential energy using the Taylor series. Similarly, this problem is not faced with weakly interacting gases because their small interactions mean that perturbation theory is applicable, despite the absence of fixed reference points. Landau summarizes this situation as liquids have no small parameter [2]. In view of this, it was believed for a long time that it was impossible to calculate liquid energy in a general form, unlike for gases and solids [1,3].

Notwithstanding these theoretical difficulties, a theoretical model that departs from the traditional perspective (see, e.g., Refs. [4,5]) for calculating liquid energy has been developed in recent years [6–8]. This model, which is known as the phonon theory of liquids, is based on Frenkel's microscopic view of a liquid, which supports solid-like longitudinal and transverse modes with frequencies larger than a characteristic frequency [9]. This theory, which covers the quantum and classical regimes and also explains the experimental decrease in liquid heat capacity from $3k_B$ to $2k_B$, was tested by Bolmatov, Brazhkin and Trachenko for the heat capacities of more than 20 different systems, with good agreement between the experimental findings and theoretical predictions [6–8]. Proctor has also conducted a "detailed and rigorous" verification of the phonon theory of liquids [10,11]. Moreover, the theory has also found extensive use as a theoretical basis for both applied and fundamental research [12–14].

One of the limitations of this theory is its reliance on the virial theorem to calculate liquid energy. For this reason, we recently proposed an alternative method of calculating liquid energy that took into account the numbers of oscillating atoms and diffusing atoms, rather than using the virial theorem [15]; our theoretical predictions were compared with the experimental specific heat of liquid mercury for both the classical and quantum cases, with good agreement being found between them [15].

The experimental liquid heat capacity per atom at constant volume decreases from $\sim 3k_B$ at low temperature to $\sim 2k_B$ at high temperature [16]. Both the phonon theory of liquids and our version can explain this universal behavior of the liquid heat capacity. However, the liquid heat capacity depends on pressure as well as temperature, and increases with pressure [17]. Although both the phonon theory of liquids and our version explain how heat capacity depends on temperature, they do not directly account for the pressure dependence of heat capacity; instead, the pressure dependence enters the theory indirectly through experimental viscosity data [6]. The liquid viscosity, $\eta$, is related to the Frenkel frequency, $\omega_F(T)$, as $\omega_F = \frac{2\pi G_\infty}{\eta}$, where $G_\infty$ is the high-frequency shear modulus, and to obtain the viscosity, the Vogel-Fulcher-Tammann (VFT) expression, $\eta = \eta_0 \exp\left(\frac{A}{T-T_0}\right)$, is generally used [6]. This is done by fitting the VFT expression to experimental viscosity data. These experimental viscosity data could contain information about the interactions in addition to information about pressure, and this implicitly makes the theory system-specific and less universal.

To the best of our knowledge, this article represents the first attempt to calculate $\omega_F$ analytically as a function of temperature and pressure using simple statistical mechanical methods. In this study, we introduce a method of evaluating the Frenkel frequency as a function of temperature and pressure by applying the concept of chemical potential, under the assumption of "a diffusive equilibrium". Based on the Frenkel frequency calculated in this way, we formulate an analytical expression for the liquid energy (or heat capacity) that is dependent on temperature and pressure, in agreement with the phonon theory of liquids. We test our approach by comparing the calculated results for the liquid heat capacity with experimental data on four noble liquids, drawn from the work of Bolmatov et al. [12], and obtain good agreement. Using these results, we propose new analytical expressions describing the Frenkel line and viscosity as a function of pressure and temperature.

## II. ANALYTICAL EVALUATION OF THE FRENKEL FREQUENCY

We begin our discussion with a review of the phonon theory of liquids, which was originally proposed by Trachenko [6] and was later extended by Bolmatov, Brazhkin and Trachenko to the quantum regime [8], and give an overview of our recent work [15]. A liquid supports one longitudinal mode of vibration and two transverse modes, with frequency $\omega > \omega_F$. In addition to this oscillatory motion, atoms in the liquid undergo diffusive motion between neighboring positions. Based on these two types of particle motion, i.e., oscillatory and diffusive motion, an expression for the liquid energy can be obtained for the anharmonic classical case as

$$E = \left(1 + \frac{\alpha T}{2}\right) N \left(3 - \left(\frac{\omega_F}{\omega_D}\right)^3\right) T \qquad (1)$$

($k_B = 1$) and for the anharmonic quantum case as

$$E = \left(1 + \frac{\alpha T}{2}\right) N \left(3D\left(\frac{\hbar \omega_D}{T}\right) - \left(\frac{\omega_F}{\omega_D}\right)^3 D\left(\frac{\hbar \omega_F}{T}\right)\right) T \qquad (2)$$

, where $D(x)$ is the Debye function, $\omega_D$ is the Debye frequency, $N$ is the number of particles, and $\alpha$ is the coefficient of thermal expansion, which allows the expression to take into account the effect of thermal expansion [7,8]. However, this model uses the virial theorem to calculate the liquid energy. Instead of following this approach, we recently introduced an alternative version of this theory in which the focus is on the numbers of oscillating and diffusing atoms, with an expression for the anharmonic classical energy as follows [15]:

$$E = \left(1 + \frac{\alpha T}{2}\right) N \left(3 - 2\left(\frac{\omega_F}{\omega_D}\right)^3\right) T + N \left(\frac{\omega_F}{\omega_D}\right)^3 T \qquad (3)$$

and where the anharmonic quantum energy is

$$E = \left(1 + \frac{\alpha T}{2}\right) N \left(3D\left(\frac{\hbar \omega_D}{T}\right) - 2\left(\frac{\omega_F}{\omega_D}\right)^3 D\left(\frac{\hbar \omega_F}{T}\right)\right) T + N \left(\frac{\omega_F}{\omega_D}\right)^3 T \qquad (4)$$

In our approach, the diffusing atoms in the liquid do not contribute to thermal expansion, unlike in the phonon theory of liquids. In both of these approaches, experimental viscosity data are used to calculate the Frenkel frequency $\omega_F = \frac{2\pi G_\infty}{\eta}$. This is a practical method that is consistent with the experimental data, but which does not explicitly take into account the pressure dependence of the heat capacity.

The factors in the expression in Eq. (3), $\left(1 + \frac{\alpha T}{2}\right)$, $N\left(3 - 2\left(\frac{\omega_F}{\omega_D}\right)^3\right)$, and $T$, are related to the thermal expansion, the number of degrees of freedom, and the mean potential or kinetic energy, respectively. A similar correspondence can be considered for Eqs. (1), (2) and (4). Moreover, the above expressions for liquid energy depend on the ratio $\frac{\omega_F}{\omega_D}$, the value of which varies from zero to one with temperature and represents how close the liquid is to either a solid or gaseous state. In other words, this ratio is closely related to the relative weights of the diffusive and oscillatory components, and controls the change in the dynamics of the particles with temperature. Hence, if we extend the model to the case where the ratio is explicitly dependent on pressure and temperature, the liquid energy or heat capacity can be calculated as a function of pressure and temperature. This can be done by taking into account an equilibrium state.

In recent work [15], we showed that the energy of a system consisting of

$$N_v = \frac{N}{3} + \frac{2N}{3}\left(1 - \left(\frac{\omega_F}{\omega_D}\right)^3\right) \qquad (5)$$

particles exhibiting solid-like oscillatory motion and

$$N_d = \frac{2N}{3}\left(\frac{\omega_F}{\omega_D}\right)^3 \qquad (6)$$

particles exhibiting diffusive gas-like motion is equal to the liquid energy. We suggest that the second term in Eq. (5) represents the number of particles undergoing transverse modes of oscillation:

$$N_s = \frac{2N}{3}\left(1 - \left(\frac{\omega_F}{\omega_D}\right)^3\right) \qquad (7)$$

A decrease in the number of the particles undergoing transverse modes of vibration causes an increase in the number of particles exhibiting diffusive gas-like motion. Since particles can move freely between pure dynamical states, an equilibrium is reached when their chemical potentials (or the fugacity) are equal. The ratio $\frac{\omega_F}{\omega_D}$ can then be determined from the relationship between the chemical potential and the number of particles. The use of the chemical potential is a good way of determining the Frenkel frequency, since it is a function of temperature and pressure, thus providing a connection between the Frenkel frequency and pressure.

In terms of the fugacity, $z = e^{\mu/k_B T}$, the average number of particles in a system is given by

$$\bar{N} = z\frac{\partial \ln \mathcal{Z}}{\partial z} \tag{8}$$

, where $\mu$ is the chemical potential, and $\mathcal{Z}$ is the grand partition function, which we can write as follows [18]:

$$\mathcal{Z} = \sum_{N=0}^{\infty} z^N Q_N(V,T) \tag{9}$$

Here, $Q_N(V,T)$ is the partition function. Using the above expressions, the chemical potentials for solid-like and gas-like motion can be calculated as follows.

We apply the indistinguishable particle approximation to describe gas-like motion in a liquid, since the particles are non-localized. The partition function for such a system can be written as

$$Q_N(V,T) = \frac{q_d(V,T)^N}{N!} \tag{10}$$

, where $q_d(V,T)$ is the partition function for a single diffusing particle. By substituting Eq. (10) into Eq. (9), we obtain the following expression for the grand partition function of diffusing atoms:

$$\mathcal{Z}_d = \sum_{N=0}^{\infty} \frac{(z_d q_d(V,T))^N}{N!} = \exp(z_d q_d(V,T)) \tag{11}$$

whereas substituting Eq. (11) into Eq. (8) gives the fugacity for diffusing atoms as follows:

$$z_d = \frac{N_d}{q_d} \tag{12}$$

The fugacity for the oscillating atoms can be calculated in the same way, except that this time we assume that the oscillators are distinguishable, because they are localized. The partition function for the oscillating atoms can be written as

$$Q_N(V,T) = q_s(V,T)^N \tag{13}$$

where $q_s(V,T)$ is the partition function for a single oscillating particle. Substituting Eq. (13) into Eq. (9) gives

$$Z_s = \sum_{N=0}^{\infty} (z_s q_s(V,T))^N = \frac{1}{1 - z_s q_s(V,T)} \tag{14}$$

and substituting Eq. (14) into Eq. (8) gives the fugacity for the oscillating atoms as

$$z_s = \frac{N_s}{q_s(N_s + 1)} \tag{15}$$

We assume that the oscillating atoms in a liquid are in equilibrium with the diffusing atoms when $z_s = z_d$ or

$$\frac{N_d}{q_d} = \frac{N_s}{q_s(N_s + 1)} \tag{16}$$

Note that we obtain a relation between the numbers of particles and their partition functions, and the numbers of particles in Eq. (16) are a function of the Frenkel frequency. We can therefore calculate the Frenkel frequency in terms of $q_d$ and $q_s$. However, in the following, we analyze a solution devised for the classical case.

If the diffusing atoms act like a classical ideal gas, then we have $q_d = V(\frac{2\pi m k_B T}{h^2})^{3/2}$ and $q_s = (\frac{k_B T}{\hbar \omega} e^{\frac{\epsilon}{k_B T}})^2$ for a two-dimensional harmonic oscillator. Here, $m$ is the mass of the atom, $V$ is the volume, $\omega$ is frequency of transverse vibration, $\epsilon$ is the binding energy of an atom in the ground state, and $h$ is the Planck constant. Substituting Eqs. (6), (7), $q_d$, and $q_s$ into Eq. (16) and applying the ideal gas law to eliminate $V$ yields

$$fx = \frac{2N(1-x)/3}{2N(1-x)/3 + 1} \tag{17}$$

, where $x \equiv \left(\frac{\omega_F}{\omega_D}\right)^3$,

$$f(P,T) \equiv \frac{P e^{\frac{2\epsilon}{k_B T}}}{\beta(m,\omega)\sqrt{T}} \tag{18}$$

$$\beta(m,\omega) \equiv \frac{3}{2}\left(\left(\frac{m}{2\pi}\right)^{3/2} \frac{\sqrt{k_B}}{\hbar} \omega^2\right) \tag{19}$$

and $P$ is the pressure. In the limit as $N \to \infty$, the solutions to the equation are given by

$$x = \frac{1 + f \mp |f - 1|}{2f} \tag{20}$$

For $f > 1$, the solutions are $x_+ = 1$ and $x_- = 1/f$, meaning that $x$ falls between zero and one, as expected. Thus, the Frenkel frequency depending on both pressure and temperature is calculated as

$$\left(\frac{\omega_F}{\omega_D}\right)^3 = \frac{\beta(m,\omega)\sqrt{T}}{Pe^{\frac{2\epsilon}{k_BT}}} \tag{21}$$

Note that the Frenkel frequency does not depend on experimental viscosity data, and that all quantities are determined by the characteristic properties of the system. By substituting Eq. (21) into Eq. (3), we can obtain an analytical expression for the energy that depends on temperature and pressure. With a few minor modifications, this approach can also be applied to the quantum case. When developing this approach, we drew inspiration from the vapor–solid equilibrium model [19].

### III. COMPARISON WITH EXPERIMENTAL DATA

In the previous section, we obtained an expression for the classical liquid energy as a function of pressure and temperature by calculating the Frenkel frequency. In this section, we compare our theoretical predictions with experimental data on the heat capacity. To enable a direct comparison with the phonon theory of liquids, we selected the four monoatomic noble liquids Ar, Ne, Kr, and Xe that were considered in Ref. [8], and compared each of them (except Kr) over the same temperature and pressure ranges in this work. Since the heat capacity for Kr is below 2 at temperatures above 470 K, we calculated the heat capacity at the same pressure as used in the previous study, but in the range 200−470 K rather than at higher temperatures as in Ref. [8]. Experimental data on the heat capacities for various temperature and pressure ranges were drawn from the National Institute of Standards and Technology (NIST) database [20].

By applying the ideal gas law and the relation $\omega = 2\pi/\tau$, we transform the function $x(T,P,\omega)$ into $x(T,V,\tau)$. We then substitute this into Eq. (3) and calculate the heat capacity $c_V = \frac{1}{N}\left(\frac{\partial E}{\partial T}\right)_V$ as

$$c_V = 3(1+\alpha T) - \left(1 + 3\alpha T + \frac{4\epsilon}{k_B}\left(\alpha + \frac{1}{T}\right)\right)\frac{x(T,P)}{2} \tag{22}$$

Calculations were also performed in the range $f(P,T) > 1$. The calculated and experimental results for the heat capacity as a function of temperature at constant pressure are presented in Figs. 1–4.

The proposed model itself contains no fitting parameters; however, since some liquid values are not precisely known, we utilized $\alpha, \epsilon$, and $\tau$ as fitting parameters, and the values used for the calculations appear to be physically reasonable.

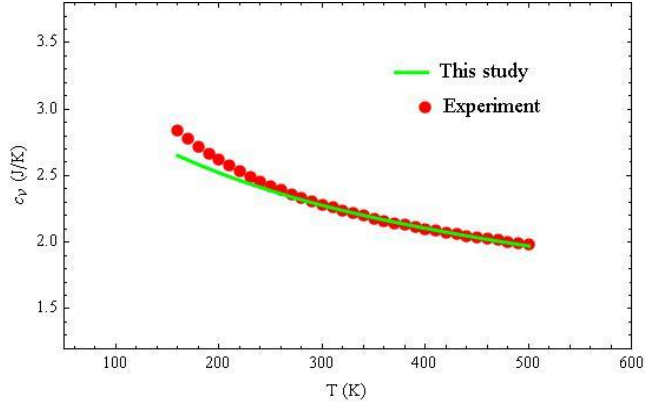

**FIG. 1:** Comparison of experimental values of the liquid heat capacity for $Ar$ with those calculated using the proposed approach. The experimental heat capacity data at $378\ MPa$ are drawn from the NIST database [20], and the value of $m$ is also taken from NIST [21]. The values used in the calculation are $m = 39.948\frac{g}{Mol}$, $P = 378$ MPa, $\epsilon = 12.39 \times 10^{-3}\ eV$, $\tau = 8.715\ ps$ and $\alpha = 8.0 \times 10^{-5}\ K^{-1}$.

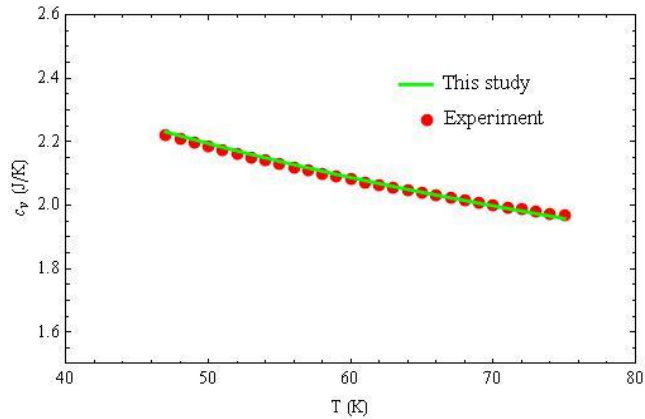

**FIG. 2:** Comparison of experimental values of the liquid heat capacity for $Ne$ with those calculated using the proposed approach. The experimental heat capacity data at 70 MPa are drawn from the NIST database [20], and the value of $m$ is also taken from NIST [21]. The values used in the calculation are $m = 20.1796\frac{g}{Mol}$, $P = 70$ MPa, $\epsilon = 1.84 \times 10^{-3}\ eV$, $\tau = 7.583\ ps$ and $\alpha = 2.2 \times 10^{-4}\ K^{-1}$. The experimental value for $\alpha$ obtained from the NIST database is $7.7 \times 10^{-3}\ K^{-1}$ [8].

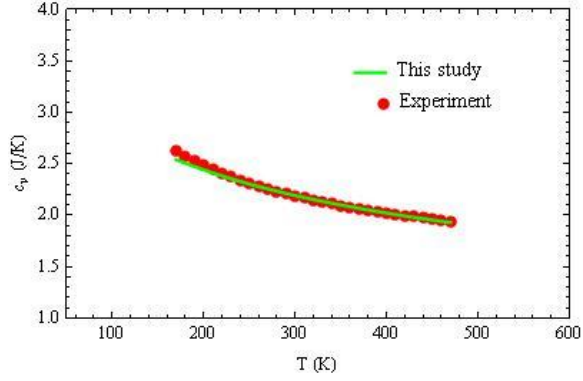

**FIG. 3:** Comparison of experimental values of the liquid heat capacity for $Kr$ with those calculated using the proposed approach. The experimental heat capacity data at 200 MPa are drawn from the NIST database [20], and the value of $m$ is also taken from NIST [21]. The values used in the calculation are $m = 83.798 \frac{g}{Mol}$, $P = 200$ MPa, $\epsilon = 1.15 \times 10^{-2}$ $eV$, $\tau = 20.52\ ps$ and $\alpha = 0$. The experimental value for $\alpha$ obtained from the NIST database is $3.6 \times 10^{-4}\ K^{-1}$ [8].

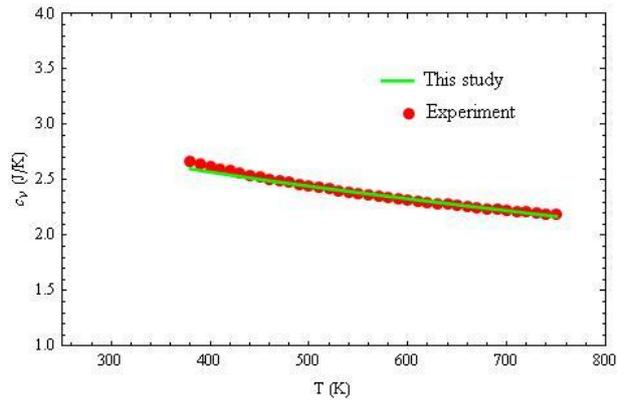

**FIG. 4:** Comparison of experimental values of the liquid heat capacity for $Xe$ with those calculated using the proposed approach. The experimental heat capacity data at 700 MPa are drawn from the NIST database [20], and the value of $m$ is also taken from NIST [21]. The values used in the calculation are $m = 131.293 \frac{g}{Mol}$, $P = 700$ MPa, $\epsilon = 1.88 \times 10^{-2}\ eV$, $\tau = 17.191\ ps$ and $\alpha = 3.6 \times 10^{-4}\ K^{-1}$. The experimental value of $\alpha$ obtained from the NIST database is $4.1 \times 10^{-4}\ K^{-1}$ [8].

## IV. DISCUSSION

In Section II, we obtained the Frenkel frequency as a function of pressure and temperature based on the condition for the diffusive equilibrium, and by combining this with Eq. (3), we derived a relationship between heat capacity and pressure. In addition to incorporating pressure, the proposed approach for calculating the Frenkel frequency remains consistent with the phonon theory of liquids. For instance, at low temperature, Eq. (21) gives $\left(\frac{\omega_F}{\omega_D}\right)^3 \approx 0$, and substituting this into Eq. (3) gives a value for the heat capacity of $c_V \approx 3$. At high temperature, Eq. (21) yields $\left(\frac{\omega_F}{\omega_D}\right)^3 \approx 1$ (since $f \geq 1$), and when this is used in Eq. (3), the heat capacity is obtained as $c_V \approx 2$. Thus, the universal behavior of the heat capacity of liquids is explained.

Recently, researchers have introduced the Frenkel line (FL) as a new dynamic boundary within the supercritical region of the phase diagram [22–25]. Below the FL, particles display both oscillatory and diffusive motion, thereby preserving rigidity and allowing high-frequency transverse modes to propagate. In contrast, above the FL, the motion is entirely diffusive, causing rigidity to vanish and making transverse modes unsustainable at any frequency. When the FL is crossed, the properties of the system change radically. The FL can be rigorously determined based on the velocity auto-correlation function [2]. However, our approach can be used as a new method to determine the FL, as follows. At the FL, transverse modes disappear, and only the longitudinal mode remains. The loss of two transverse modes corresponds to $\left(\frac{\omega_F}{\omega_D}\right)^3 \approx 1$, or according to Eq. (20), $f(P,T) = 1$. Thus, we can obtain an analytical expression that determines the FL at any pressure and temperature.

The value of $\eta$ exhibits strong system dependency, and is also affected by changes in temperature and pressure [17]. It is determined by the activation energy barrier $U$ for molecular rearrangements, which is itself linked to the nature of interatomic interactions and the structure of the system. Due to the complexity of these factors, there is no universal method for predicting $U$ and $\eta$ using first-principles theories. For this reason, as discussed above, the Frenkel frequency is derived from experimental viscosity data using $\omega_F = \frac{2\pi G_\infty}{\eta}$ and the VFT expression [6]. However, our approach could be used as a toy model for studying viscosity, since we calculate the Frenkel frequency as a function of pressure and temperature in Eq. (21), and the Frenkel frequency can be written in terms of liquid viscosity $\eta$ using the relationship $\omega_F = \frac{2\pi G_\infty}{\eta}$.

Moreover, as the temperature decreases, the viscosity increases, and the system approaches the solid state and begins to behave like a solid. In other words, at low temperatures, most of the particles are localized and distinguishable. Due to the localized nature of the oscillating particles, we did not include the factor $N!$ in Eq. (13), unlike in Eq. (10). Since this factor is used in the calculation of the Frenkel frequency, it is also directly related to viscosity. This relationship can be interpreted as follows: as the quantity of localized (or distinguishable) particles grows, the viscosity undergoes a corresponding increase.

In the calculation of the Frenkel frequency, the solution to Eq. (20) for $f(T,P) > 1$ was used to ensure that $x(T,P)$ was between zero and one. However, Eq. (20) also has a solution for $f(T,P) < 1$, where in this case, $x(T,P) > 1$. These distinct solutions may potentially correspond to two regions in the phonon theory of liquids [2]: the hydrodynamic region ($\omega\tau < 1$; for the definitions of $\omega$ and $\tau$, see Ref. [2]) and the solid-like elastic region ($\omega\tau > 1$).

Note that in the classical case, the liquid heat capacity is independent of the Debye frequency $\omega_D$. In other words, in the classical case, the calculation of the liquid energy does not require the Debye frequency $\omega_D$. However, in the quantum case, due to the Debye functions in Eqs. (2) and (4), $\omega_D$ is necessary to determine the liquid energy.

We have limited our discussion here to the classical case where all oscillators have the same frequency $\omega$, and have focused on understanding the physical mechanism of the microscopic dynamics in terms of temperature and pressure. However, our approach remains adaptable, and can also be applied to the quantum case.

## V. CONCLUSION

In this article, we have proposed a new model for calculating the Frenkel frequency in terms of pressure and temperature. Based on this calculated Frenkel frequency, we derived an analytical expression for the liquid heat capacity as a function of temperature and pressure. To test the proposed approach, we compared our theoretical predictions with experimental data for four noble liquids from the work in Ref. [8], and found good agreement. The results from the proposed approach were used to offer suggestions for determining the FL and evaluating viscosity.